\documentclass[journal]{IEEEtran}
\IEEEoverridecommandlockouts
\usepackage{cite}
\usepackage{comment}
\usepackage{amsmath,amssymb,amsfonts}
\usepackage{algorithmic}
\usepackage{graphicx}
\usepackage{textcomp}
\usepackage{caption}
\usepackage{subcaption}
\usepackage{upgreek}
\usepackage{url}
\usepackage{multirow}
\usepackage{color}
\usepackage[normalem]{ulem}
\usepackage{xcolor}
\usepackage[labelfont={color=black}]{caption}

\newcommand\rev[1]{\textcolor{black}{#1}}

\newcommand\revSecond[1]{\textcolor{black}{#1}}
\newcommand\revThird[1]{\textcolor{black}{#1}}

\def\BibTeX{{\rm B\kern-.05em{\sc i\kern-.025em b}\kern-.08em
    T\kern-.1667em\lower.7ex\hbox{E}\kern-.125emX}}
\begin{document}

\title{\textbf{SERAD}: \underline{S}oft \underline{E}rror \underline{R}esilient \underline{A}synchronous \underline{D}esign using \rev{a} Bundled Data Protocol}

\author{\IEEEauthorblockN{Sai Aparna Aketi\IEEEauthorrefmark{2}, Smriti Gupta\IEEEauthorrefmark{2}, Huimei Cheng\IEEEauthorrefmark{1}, Joycee Mekie\IEEEauthorrefmark{2} and Peter A. Beerel\IEEEauthorrefmark{1}\IEEEauthorrefmark{3}}
\thanks{\IEEEauthorrefmark{2}S. A. Aketi, S. Gupta, and J. Mekie are with the Department of Electrical Engineering, Indian Institute of Technology Gandhinagar (IITGN), Gandhinagar, Gujarat 382355,  India (email: saketi@purdue.edu; smriti.gupta@mtech2016.iitgn.ac.in; joycee@iitgn.ac.in)}
\thanks{\IEEEauthorrefmark{1}H. Cheng and P. A. Beerel are with the Ming Hsieh Dept. of Elec. and Comp. Eng., University of Southern California, Los Angeles, CA. USA (email: huimeich@usc.edu; pabeerel@usc.edu)}
\thanks{\IEEEauthorrefmark{3}P. A. Beerel also consults for Galois, Inc in the area of asynchronous design.}
}

\maketitle

\begin{abstract}
The risk of soft errors due to radiation continues to be a significant challenge for engineers trying to build systems that can 
handle harsh environments. 
Building systems that are \revSecond{Radiation Hardened by Design} (RHBD) is the preferred approach, but existing techniques are 
expensive in terms of performance, power, and/or area.
This paper introduces a novel soft-error resilient asynchronous bundled-data design template, \emph{SERAD}, which uses a combination of temporal and spatial redundancy to mitigate \revSecond{Single Event Transients} (SETs) and upsets (SEUs). SERAD uses \revSecond{Error Detecting Logic} (EDL) to detect SETs at the inputs of sequential elements and correct them via re-sampling. Because SERAD only pays the delay penalty in the presence of an SET, which rarely occurs, 
its average performance is comparable to the baseline synchronous design. 
We tested the SERAD design using a combination of Spice and Verilog simulations and evaluated its impact on area, frequency, and power on an open-core MIPS-like \rev{processor} using a NCSU 45nm cell library. Our post-synthesis results show that the SERAD design consumes less than half of the area of the \revSecond{Triple Modular Redundancy}  (TMR), exhibits significantly less performance degradation than \revSecond{Glitch Filtering} (GF), and consumes no more \rev{total} power than the baseline unhardened design.

\end{abstract}


\section{Introduction}

Designing electronic components that can sustain harsh environmental conditions in space and military applications has been an important field of research for many decades \cite{Gouker13,Wong11}. Prolonged bombardment of heavy-ions, protons, neutrons and other particles on electronic equipment and circuits can result in permanent damage or soft errors that cause temporary failures \cite{Hopkinson:00,Mahatme:15}. When high-energy neutrons (present in terrestrial cosmic radiations) or alpha particles (that originate from impurities in the packaging materials) strike a sensitive node in a CMOS circuit, they generate a dense local track of additional electron-hole pairs in the substrate. This additional charge is collected by the drain of an OFF transistor and can result in a transient voltage pulse \cite{Hosseinabady:2006}.


Because these events are somewhat rare, typical models assume these events can be analyzed in isolation as a \revSecond{Single Event Transient} (SET), but as technology scales multiple event transients are also becoming relevant. 
A transient event can be altered as it propagates along a path through combinational circuit. 
In particular, an event may be masked due to three techniques, i.e. logical masking, electrical masking, and temporal masking. In addition, the event may be attenuated or propagated and can result in an early or late edge or a dynamic hazard \cite{Hosseinabady:2006}. When the transient is finally latched, a \revSecond{Single Event Upset} (SEU) is created.  Traditionally transients that effect memory elements were considered more significant than those that strike combinational logic, but as technology has scaled, strikes in the combinational logic are also becoming important~\cite{Wong2014}.

To circumvent the effects of these types of radiation, numerous \revSecond{Radiation Hardened by Design} (RHBD) techniques have been proposed in the literature, the most common being the \revSecond{Triple Modular Redundancy} (TMR)~\cite{TMR}, \revSecond{Guarded Dual Modular Redundancy} (GDMR)~\cite{Mekie2016}, gate sizing~\cite{Zhou:2006,Rao2006,Zivanov06,Sabet2017},  and \revSecond{Glitch Filtering} (GF)~\cite{Bhuva03,Naseer:2005,bala:2005}. 
RHBD cell libraries have also been developed for both asynchronous and
synchronous designs~ \cite{Calligaro:2008,David:2009}. One study demonstrated the benefits of using asynchronous bundled-data latch-based design techniques 
to reduce the area, delay, and power penalty associated with a RHBD cell library ~\cite{David:2009}. However, radiation-hardened libraries are cumbersome to design, typically generations behind state-of-the-art unhardened versions, and relatively expensive in terms to cell area, power consumption, and cell delay. 

This paper explores the possibility of building an efficient SET-resilient technique that uses standard cell libraries and pays a performance penalty only when \revSecond{an} SET occurs. The proposed approach is inspired by an asynchronous bundled-data template that is timing-resilient~\cite{Peter:2015}, exhibiting high performance when no timing errors occur and gracefully slowing down in the presence of timing errors. 
However, unlike timing-resilient designs that can assume that the timing of signals is governed by a notion of worst-case 
delay, SET-resilient circuits must account for the fact that SETs can occur at any time. This means the existing timing resilient design cannot be simply adopted and new SET-aware asynchronous protocols and circuits must be developed.

Our proposed template that achieves this goal is called SERAD which refers to \emph{s}oft \emph{e}rror \emph{r}esilient \emph{a}synchronous \emph{d}esign using a bundled data protocol. 
Like its timing-resilient cousin, it uses standard single-rail combinational logic and error-detecting latches~\cite{Bowman:2009}. However, its handshaking protocol and control logic are completely redesigned to use a combination of temporal and spatial redundancy to mitigate SETs. 
The use of single-rail combinational logic yields an area and power efficient design, and the novel error-detecting logic and control structure makes it performance efficient as it delays the pipeline 
only when \revSecond{an} SET occurs. 
Given that soft-errors are quite rare~\cite{anand:2011}, the average performance of SERAD is thus essentially the same as its non-SET-resilient counterpart.  

\begin{figure*}[htb]
        \centering
        \vspace*{2em}
        \begin{subfigure}[b]{0.5\textwidth}
                \centering
                \includegraphics[width=0.8\textwidth]{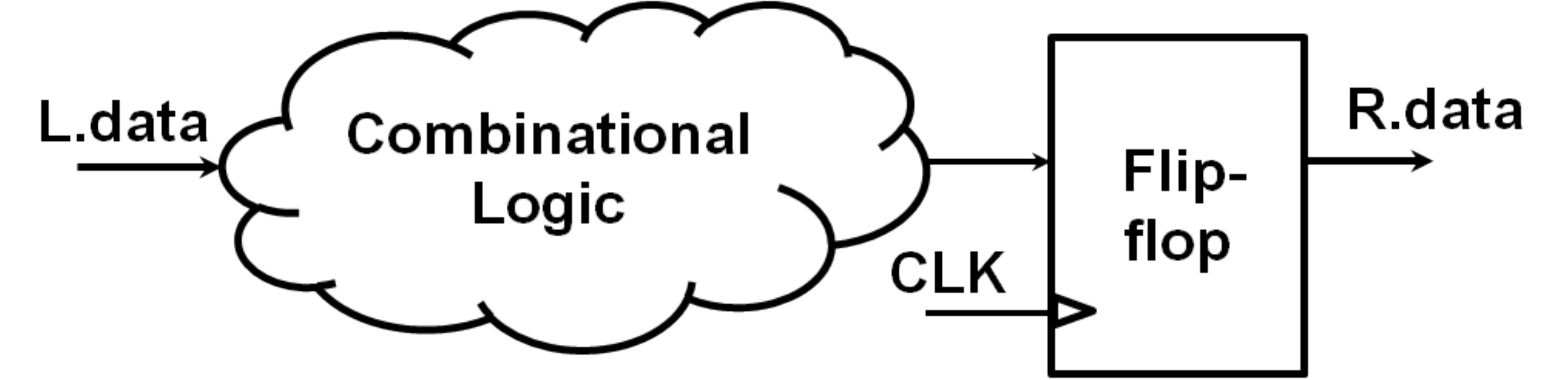}
                \caption{Synchronous template}
                \label{fig:sync template}
        \end{subfigure}\\
        \vspace*{1em}
         \begin{subfigure}[b]{0.45\textwidth}
                \centering
                \includegraphics[width=\textwidth]{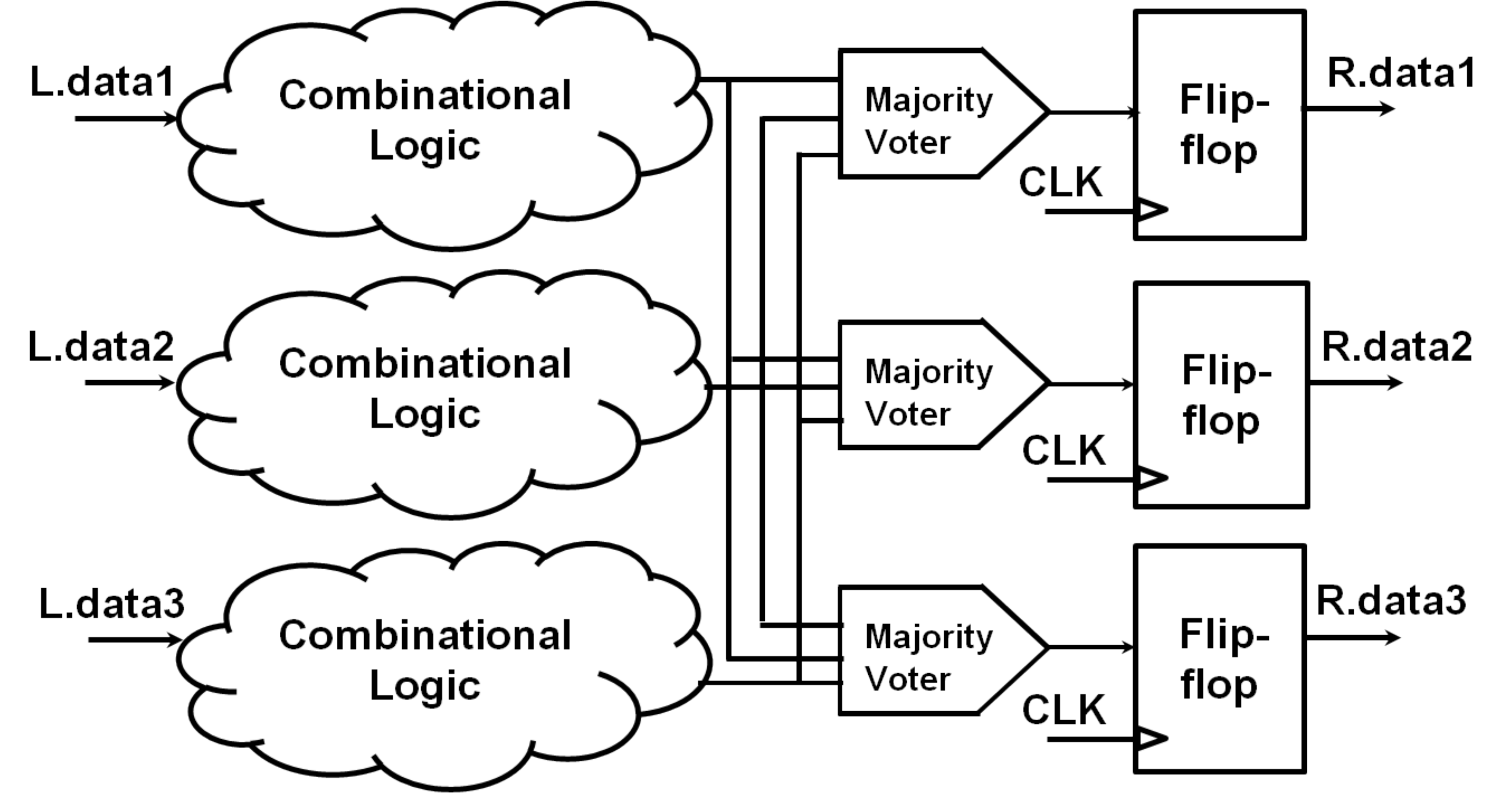}
                \caption{TMR template}
                \label{fig:tmr template}
        \end{subfigure}%
        \begin{subfigure}[b]{0.45\textwidth}
                \centering
                \includegraphics[width=\textwidth]{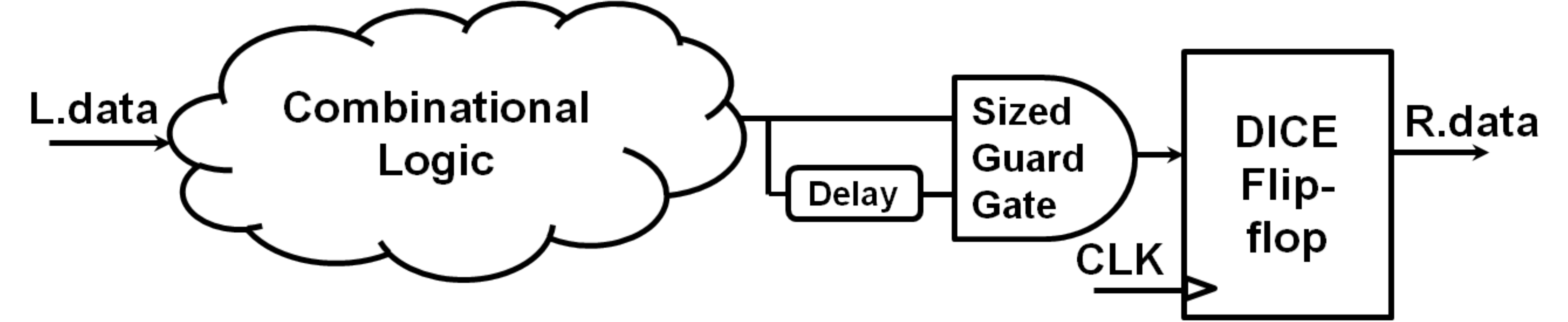}
                \caption{Glitch-filter template}
                \label{fig:gf template}
                 \end{subfigure}
         \vspace*{1em}
         
\caption{Block diagram of original flop-based circuit and two well-known SET-resilient variants.}
\label{fig:Variants}         
\end{figure*}

\bigskip
The remainder of this paper is organized as follows.
Section~\ref{sec:related} describes related research to make synchronous and asynchronous techniques resilient to soft errors. 
Section~\ref{sec:SERAD Template} then presents a detailed description of the proposed SERAD template describing how it combines spatial and temporal redundancy to ensure SET resilience.
Section~\ref{sec:exp_results} presents our experimental results that consist of analog and digital simulations that validate SERAD's SET resilience as well as a case study on an open-core MIPS-like \rev{processor} that quantifies SERAD's relative area, frequency, and power compared to alternative approaches. Finally, Section~\ref{sec:conclusions} 
summarizes the paper and describes opportunities for future work.

\section{Related Work}
\label{sec:related}

The gold standard approach to making a \rev{typical  synchronous  system} shown in Fig.~\ref{fig:Variants}(a) resilient to soft-errors 
is TMR. As illustrated in Fig.~\ref{fig:Variants}(b), this  
uses three copies of the combinational logic and \revSecond{flip-flops (FFs)}
and an additional voting structure to pick 
the answer common to at least two copies. 
When applied to a large system,
this approach is complicated by the need to re-synchronize the blocks 
after an event occurs. 
Eaton et al. addressed this issue by applying this concept to the design of
\revSecond{FFs} that sample the data at three distinct time steps, 
each separated from the next by at least the maximum \revSecond{SET} pulse width \cite{Eaton2002}.
This is reliable but has high area, performance, and power penalties.
Dual-mode and dual-rail versions use two copies of the logic followed by a 
Muller C-element that waits until both copies agree before firing 
\cite{Eaton2002,Mekie2016} and are also costly.
Alternatively, gates and storage elements can be sized to minimize the 
risk of soft-errors \cite{Zivanov06,Rao2006,Sabet2017}.
Unfortunately, the increase in size needed for the gates is technology 
dependent and expected to grow as technologies scales. 
In addition, numerous researchers have focused on making 
storage elements and memory arrays SET-tolerant, the \rev{latter} using error-detecting 
codes \cite{Calin:1996, Kobayashi:2014, Mitra2006}. 
Our template leverages these techniques as well.
Other researchers have proposed to add glitch filters to the output of combinational 
logic blocks and thereby mask SETs before they can be latched \cite{Bhuva03,Naseer:2005}, as illustrated in Fig.~\ref{fig:Variants}(c). 
However, the glitch filter effectively increases the latency of the combinational 
logic and does not combat a late edge that can be caused by \revSecond{an} SET.
Therefore to use this approach the clock period must be increased 
by at least two times that of the maximum width of an SET. Several experimental
efforts have reported glitch widths between 10s of pico-seconds to over 1ns \cite{Eaton:2004}, 
suggesting that the performance penalties can be impractical for high-speed designs.
An important point to be made is that in all these schemes the performance penalty is uniform \rev{in that it increases the critical path of the design, thereby decreasing its maximum clock frequency.} 
\rev{Therefore, for designs with little timing slack, the performance penalty will exist} in every cycle of operation independent of whether a soft-error occurs or not.

There \rev{have} also been several works that have tried to leverage asynchronous design as a basis for radiation hardening. They can be divided into two domains - one domain is based on using dual-rail (DR) or quasi-delay-insensitive (QDI) logic and the other is based on bundled-data design. The former covers works based on many existing asynchronous templates, including weak-conditioned half-buffers~\cite{Pontes:2013}, dual-rail minterms~\cite{Monnet:2005}, pre-charged half-buffers~\cite{Jang:2005}, null-convention logic~\cite{Kuang:2010,Di:2007}, and others~\cite{Friesenbichler:2009}.
In each case, the researchers add circuitry to traditional DR/QDI circuit templates to make them more resilient to soft-errors. However, the base circuitry often requires at least two times more transistors than standard single-rail synchronous logic~\cite{Proteus2011}, incurring large penalties in silicon area and leakage current when compared to standard synchronous design. Moreover, the \rev{return-to-zero} nature of these templates often implies higher switching activities than synchronous single-rail counterparts which results in higher dynamic power. While these asynchronous design styles are attractive because they require few timing assumptions to ensure correctness and can thus work in harsh environments, their costs may be prohibitive for many applications.

In contrast, bundled-data asynchronous circuits use standard combinational logic and sequential elements. They differ from their synchronous counterparts only in that instead of \revThird{a} global clock, asynchronous circuits are used to generate local clock trees. The result is a design which is similar (if not better) in area and power consumption than synchronous designs but require timing assumptions to ensure correctness. 
Delay lines are needed as part of the asynchronous circuits to ensure setup and hold constraints are satisfied. 
As these can be difficult to design optimally they are often made to be post-silicon programmable.  
The sequential elements can be latch or \revSecond{FF} based, latch-based typically being more flexible and lower power \cite{David:2009}. 
Making these designs radiation \revThird{hardened} has often focused on the unique aspects of the asynchronous nature - the asynchronous control logic~\cite{Naqvi:2014,Danilov:2016}. One of the simplest approach is dual-modular redundancy with guard-gates~\cite{Almukhaizim:2009}. This doubles the size of the control logic, but given the control logic is a small fraction of the overall chip area, the penalties are quite manageable. SETs in the combinational logic, however, must be handled through a combination of gate-sizing, spatial redundancy, or temporal redundancy, similar to other synchronous designs. 
SERAD is also based on bundled-data but is, to the best of our knowledge, the first to be inspired by a timing-resilient form of bundled-data design. 

\begin{figure}[!t]
  \centering
    \includegraphics[width=0.48\textwidth]{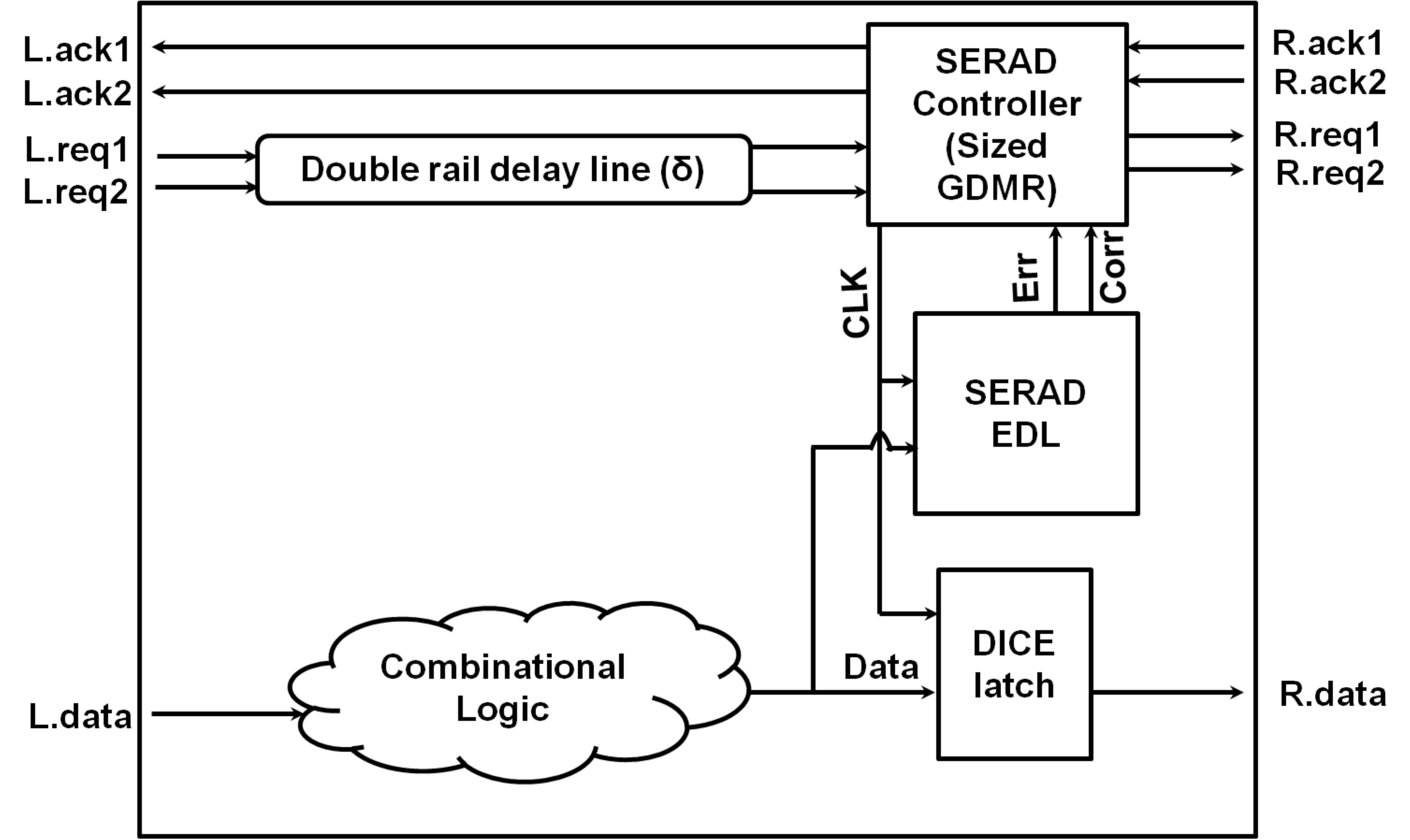}
    \caption{Block diagram of proposed SERAD template}
    \label{fig:SERAD_template}
\end{figure}

\section{SERAD Template}
\label{sec:SERAD Template}

The proposed SERAD template, illustrated in Fig.~\ref{fig:SERAD_template}, uses single-rail combinational logic,  error detecting logic (EDL), delay lines, DICE latches, and a novel SERAD controller.
%
%
SERAD mitigates SETs that originate in the sequential elements of the pipeline, i.e, latches, using DICE latches.  
SERAD prevents SETs that originate in the combinational logic from propagating from one pipeline stage to the next using a special form of temporal redundancy. 
Any SET which appears at the input of a pipeline latch when it is transparent is identified 
as an error by the EDL and is mitigated by stalling the pipeline until the data is re-sampled. 
In particular, the SERAD controllers communicate with each other using a \textit{soft error resilient handshaking protocol} 
that can ensure any SET of width less than a predefined value $\sigma$ is mitigated. 

Before explaining this protocol in detail, however, we first describe the delay notations critical to a SERAD design.
\begin{itemize}
    \item $\tau$: Maximum pulse width of \revSecond{an} SET that will be detected and mitigated. It is chosen such that the probability of occurrence of \revSecond{an} SET of width greater than $\tau$ in the given technology is negligible. 
    \item $\phi$: Minimum pulse-width required by the DICE latch to correctly sample data.
    \item $\sigma$: Maximum($\phi$,$\tau$)
    \item $\Delta$: Worst case delay of the combinational logic or critical path delay including the propagation delay through one DICE latch.
    \item $\delta$ = $\Delta - \sigma$. This defines the required delay of the delay line between controllers, as illustrated in Fig.~\ref{fig:SERAD_template}, ignoring the overhead of the EDL and control circuit which will be discussed in Section~\ref{sec:timing}. 
    \item $y$: Time required to close and re-open a DICE latch
\end{itemize}

\begin{figure}[!t]
        \centering
        \includegraphics[width=\linewidth]{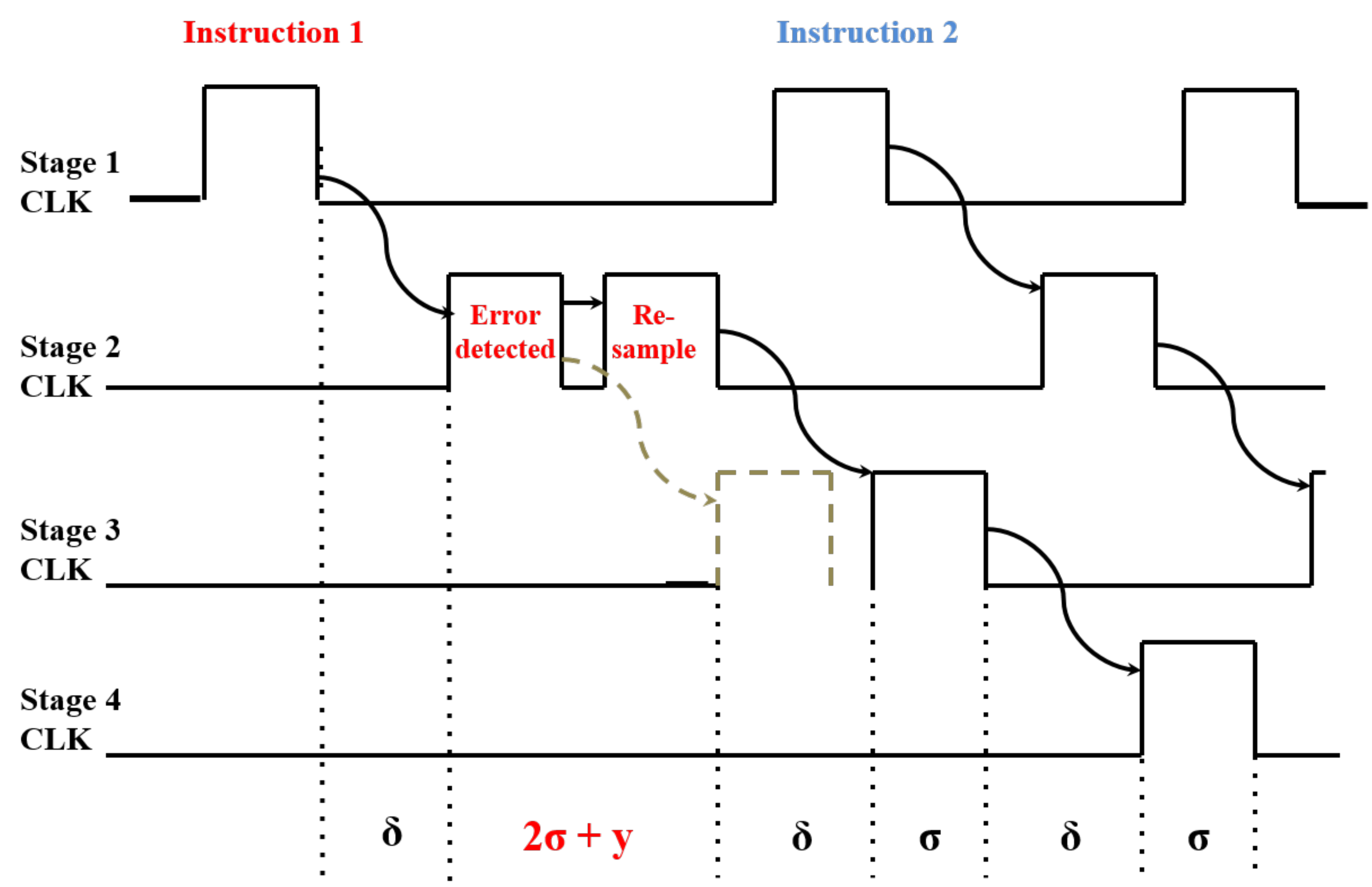}
        \caption{Timing diagram of the proposed SERAD template}
    \label{fig:timing}
    \vspace{-1em}
\end{figure}

\subsection{Soft Error Resilient Handshaking Protocol}
\label{sec:spec-handshaking}

Fig.~\ref{fig:timing} illustrates the expected behavior of the \emph{CLK} signals associated with two instructions flowing through a four-stage SERAD pipeline. 
Instruction~1 
launches from Stage 1. \revSecond{An} SET that occurs in the combinational logic path between Stage 1 and Stage 2  is detected at the Stage 2 latch. The rising edge of Stage 3's \emph{CLK} signal is nominally scheduled to occur $\delta$ time units after Stage 2's \emph{CLK} closes for the first time, shown as the dotted gray region. However, because of the occurrence of \revSecond{an} SET, the Stage 2's latch closes and re-opens, giving Instruction~1 a total of $(\delta+(2*\sigma)+y)$ time to pass from Stage 2 to Stage 3, where $y$ is the time taken to close the latch, sample the error, and reopen the latch. 
\rev{Subsequently,} instruction~2 
does not suffer a timing violation in Stage 2, which allows Stage 3's \emph{CLK} signal to rise $\delta$ time units after Stage 2's \emph{CLK} falls.

Note that if the data at the input of the latch is stable when the \emph{CLK} is high as well as during the subsequent hold time of the latch, then the latched data is error free. This interval is precisely when the EDL checks for an error and is defined as the \textit{SET Filtering Window (SFW)}. Its length is $\sigma$ plus the hold time of the DICE latch, where $\sigma$, as indicated above, is the maximum of $\phi$ and $\tau$. \revThird{This ensures} that the SFW is large enough to safely catch the SETs. The fact that the error detection window also includes the latches' hold time is unique to SET-resilient design and is motivated in Section \ref{sec:metastability}.

\subsection{SERAD Controller Design}
\label{sec:controller}

\begin{figure}[!t]
        \centering
        \includegraphics[width=\linewidth]{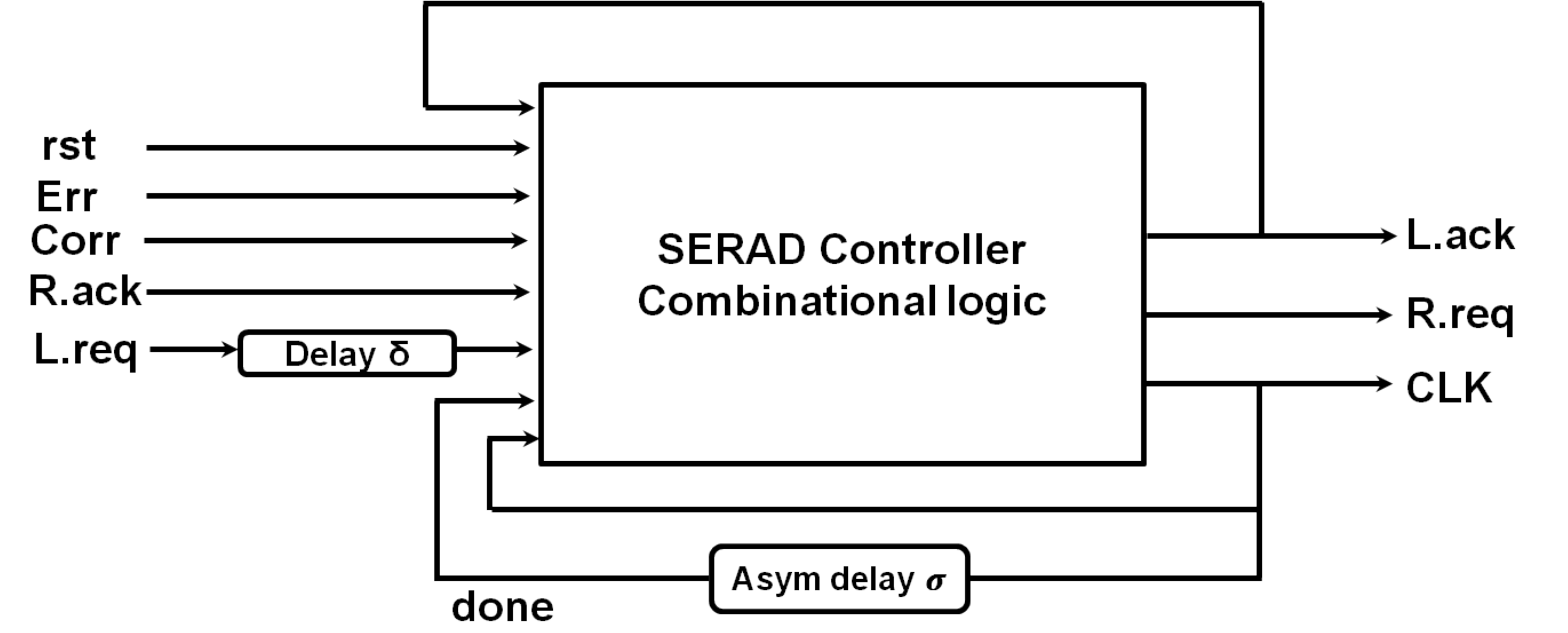}
        \caption{SERAD single-rail controller on which sized GDMR controller is based.}
    \label{fig:src}
    \vspace{-1em}
\end{figure}

\begin{figure}[!t]
        \centering
        \includegraphics[width=\linewidth]{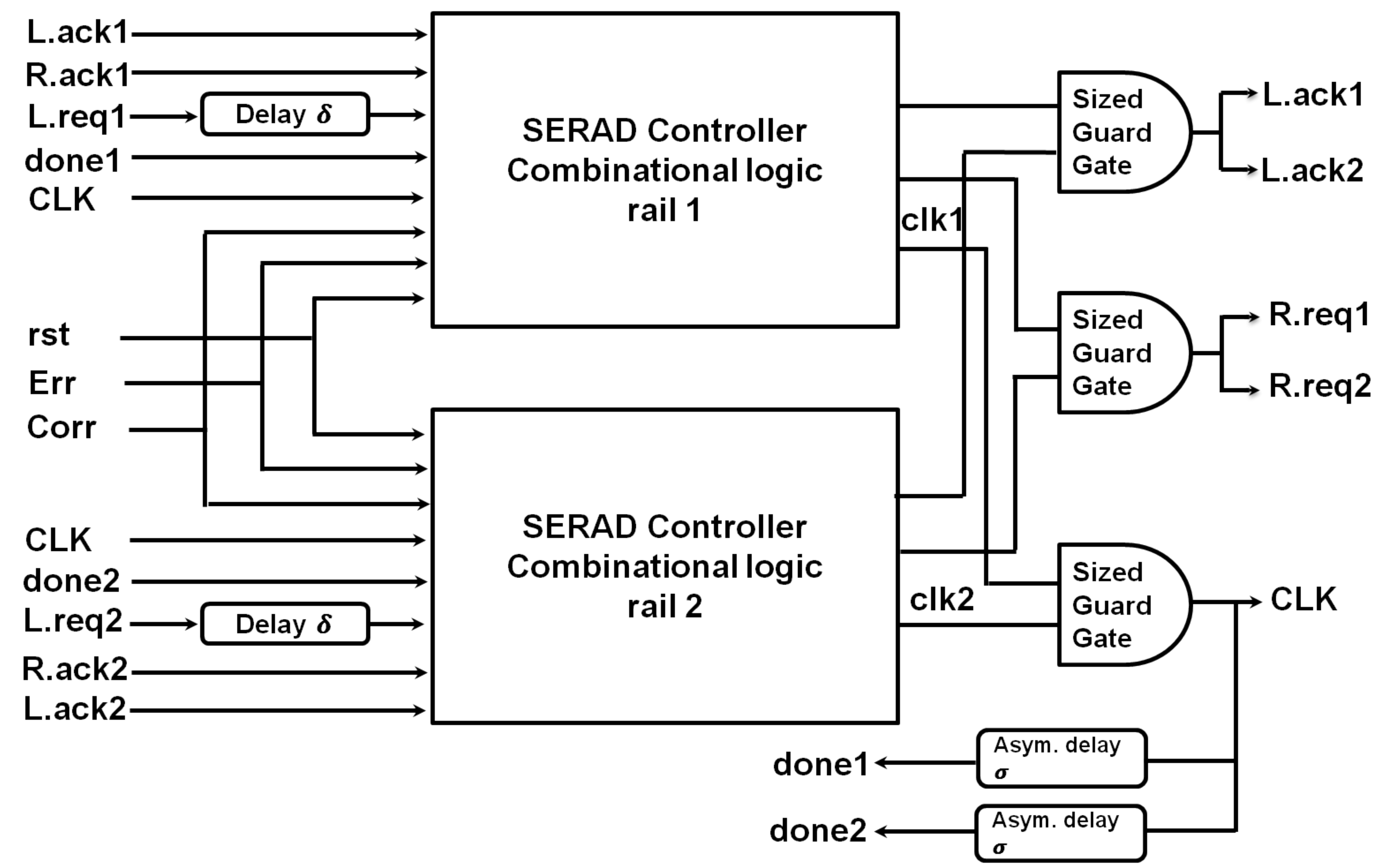}
        \caption{Sized GDMR SERAD controller}
    \label{fig:drc}
    \vspace{-1em}
\end{figure}

The SERAD controller is based on the single-rail controller shown in Fig.~\ref{fig:src} and is illustrated in Fig.~\ref{fig:drc}.  
It is designed using a combination of GDMR~\cite{Joycee:2018} and gate sizing~\cite{Zhou:2006} techniques and we call it \revThird{a} sized GDMR SERAD controller. 
The SETs on the internal nodes of the controller are mitigated via spacial redundancy and guard gates while the SETs on the output nodes are mitigated using gate-sizing. 
In particular, the feedback loops in the single-rail controller are cut and the resultant combinational logic is duplicated to introduce spatial redundancy. Each pair of redundant outputs is passed through a sized guard gate. The guard gates are sized such that any SET on the output node with a pulse width less than or equal to $\sigma$ is mitigated. Similarly, any SET of pulse width less than or equal to $\sigma$ on the internal or input nodes is filtered out by the guard gates. For this work, we have used 
\revThird{the} gate-sizing technique proposed by Zhou \emph{et al} in \cite{Zhou:2006}. In this work, the authors \revThird{calculate} the width of an inverter to remove the effect of SET \revThird{by reducing its amplitude}. For a given LET (linear energy transfer) value, the equivalent amount  of charge  ($Q$) that can \revThird{be} deposited is first computed. Needless to say, $Q$ is process dependent. The design is sized (i.e. the width is increased) to increase the node capacitance until the resulting SET pulse is sufficiently attenuated in amplitude~\cite{Zhou:2006}. We have implemented the gate-sizing algorithm proposed in ~\cite{Zhou:2006} to size the guard-gate for radiation hardening. We note here that our sized guard gate is about $10\times$ \revThird{larger than a minimum sized guard gate}. Further, sizing the guard gate also motivates the preceding circuits to be appropriately sized. We observe that the latency of the controller with sized guard gate is $6\%$ larger than the controller with minimum sized guard gates. 

\begin{figure*}[!t]
        \centering
        \includegraphics[width=0.7\linewidth]{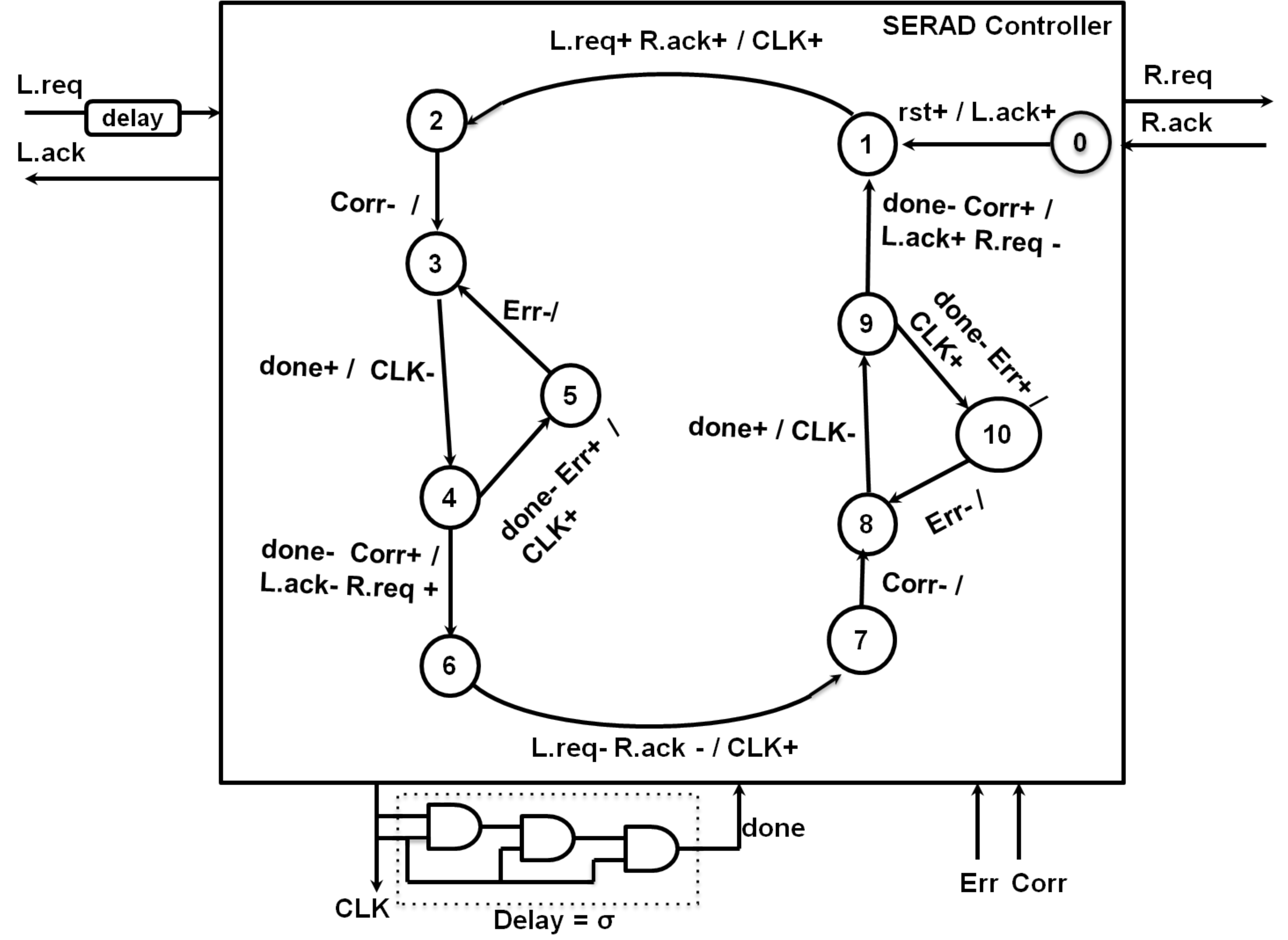}
        \caption{Burst-mode state machine of single-rail SERAD controller}
    \label{fig:bmm}
\end{figure*}

The single-rail controller is implemented as a burst-mode state machine~\cite{Hazard-free_asyncFSM:1995} and synthesized using the 3D tool~\cite{3DSynthesis:1992}.\footnote{\rev{Note that the reset signal \emph{rst} 
is added to the specification to ensure that the state machine initializes properly.}} 
%
\rev{We then implemented the GDMR SERAD controller by duplicating the single rail controllers, and combining their outputs through the sized guard gates.}  The \rev{non-reset} input signals to the first rail of the controller are \emph{
L.req1, R.ack1, Err} and \emph{Corr} and the output signals are \emph{CLK, L.ack1} and \emph{R.req1}. The input signals to the second rail of the controller are \emph{
L.req2, R.ack2, Err} and \emph{Corr} and the output signals are \emph{CLK, L.ack2} and \emph{R.req2}. The output rails are independent but the input signals to the \rev{two} rails are dependent. 
Most of the inputs are driven by the output of neighbouring controllers which are error-free as the final output nodes are sized. 
The exception is the signals \emph{Err} and \emph{Corr} that 
are driven from \rev{the} SERAD EDL which is discussed in 
Section \ref{sec:EDL}. 

Fig.~\ref{fig:bmm} shows the burst-mode state machine specification for a single-rail controller of a normal pipeline stage. The behavior of the machine can be summarized as follows.
 \begin{itemize}
     \item On deactivating reset signal, the controller sends an acknowledgement to the previous stage (moving from state 0 to state 1).
     \item On receiving a request from previous stage and acknowledgement from next stage, the controller sets the clock signal high, waits for the EDL error signal to reset, and waits for \revThird{the} done signal to go high via the illustrated asymmetric delay line (moving from \revThird{state} 1 to 2 to 3).
     \item As soon as the done signal goes high, the clock is driven low and the controller waits to receive the error information (in state 4).
     \item If \emph{Err} goes high, indicating the occurrence of \revThird{an} SET, the clock is driven high and the controller loops back to state 3.
     \item If \emph{Corr} goes high, the controller sends a request to the next stage and an acknowledgement to the previous stage, signaling that the latched data is error free (state 6).
 \end{itemize}
Due to the two-phase nature of the machine the states \revThird{1-5} are replicated in states \revThird{6-9} with the alternate signal transitions. 
Note that the controller for \revThird{other} pipeline stages \revThird{is} similar, but \revSecond{is} designed to generate an output request after reset (an initial token on reset).
\revSecond{This type of controller is called a {\em token controller} \cite{Beerel:2010} and generates} the needed initial state to the system, 
avoiding deadlock right after reset.

\rev{The synthesized logic expressions of the single-rail controller are given below.} \revSecond{The 3D synthesis tool analyzes the state space of the specification and guarantees that these expressions are free of static and dynamic hazards. In some cases, this synthesis process requires the introduction of state variables, such as the signal $z$ below. Each expression effectively describes the next state expression of one output signal that can be implemented with combinational logic but in many cases these output signals must be fedback as inputs to the controller. } 
\rev{
\begin{eqnarray*}
Rreq &=& rst .(\overline{Corr} . \overline{Lack} + Done . \overline{Lack} + z.\overline{Lack}  \\ && + z. Corr . \overline{Done}) \\
Lack &=& rst .(\overline{Corr} . Lack + Done .Lack + \overline{z}. Lack \\  & & + \overline{z}. Corr . \overline{Done} ) \\
clk &=& rst .(Err. \overline{Done} + clk. \overline{Done} + z. \overline{Lreq}. \overline{Rack} \\
 & & +~~ Lreq. Rack.  \overline{Done}.\overline{z}) \\
z &=& rst .(Done . Lack + z. Corr + \overline{z}. Lack   \\
 & & +~~ z. \overline{clk}. \overline{Err})
\end{eqnarray*}}%
\revSecond{Note that some delays were added to the controller feedback wires to avoid essential hazards. 
In addition, other delays were added to some of the 
primary outputs to avoid violations of the fundamental mode assumption \cite{Nowick:1995}}.


\begin{figure}[!t]
        \centering
        \includegraphics[width=0.98\linewidth]{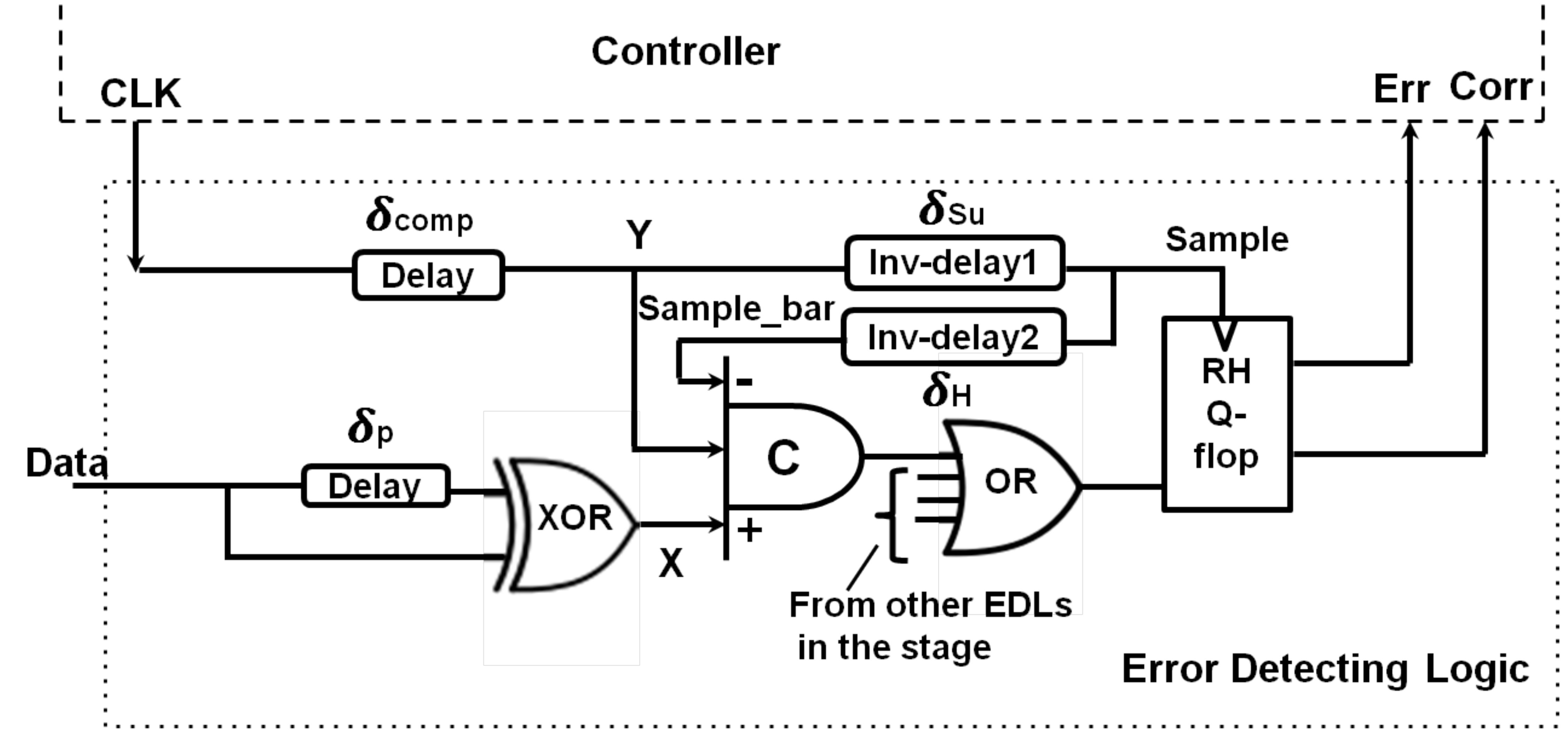}
        \caption{Block diagram of SERAD error detecting logic}
        \label{fig:edl}
    \vspace{-1em}
\end{figure}

\subsection{SERAD Error Detection Logic}
\label{sec:EDL}

As illustrated in Fig.~\ref{fig:edl}, the EDL consists of transition detectors, asymmetric C-elements, and a Q-Flop~\cite{Rosenberger:1988}. 
For our design, we implemented the EDL using Transition Detecting Time Borrowing (TDTB) latches proposed in~\cite{Bowman:2009}.
The input to the DICE latch, \emph{Data}, is fed to a transition detector consisting of an XOR gate and delay element $\delta_p$. 
The asymmetric C-element remembers errors detected by the transition detector during the high phase of 
either \emph{CLK} or \emph{Sample\_bar}. 
The \revSecond{outputs} of the C-elements are ORed together 
and sampled by a Q-flop. The Q-flop 
produces dual-rail outputs \emph{Err} and \emph{Corr} which are both initially 0.  After \emph{Sample} raises, the Q-flop samples its input. If its input is 1, i.e., if there was \revSecond{an} SET, it raises \emph{Err}; Otherwise it raises \emph{Corr}. It resets both outputs when \emph{Sample} falls. 

There are two small delay lines, typically made of a chain of inverters, that help ensure setup/hold conditions of the Q-flop are generally met.
In particular, \emph{Sample} is a sufficiently
delayed inverted version of \emph{CLK} that ensures 
that the C-element's output 
propagates to the Q-flop 
before \emph{Sample} goes high, 
thereby adhering to the Q-flop's setup time. \emph{Sample\_bar} is used as an input to the C-element
to ensure the C-element output  
resets only after the Q-flop samples the data, 
thereby adhering to the Q-flop's hold time. 
The explicit constraints
on these delays are further explained in Section \ref{sec:timing}. 

Notice that if there is \revSecond{an} SET on any of the nodes within the EDL, the EDL may produce a false-alarm that causes the controller to re-sample the latches inputs. 
It is important to recognize that this causes a degradation in performance but no logical 
error. For this reason most of the EDL 
need not \revSecond{to} have the overhead of being made SET tolerant. 
In particular, only the Q-flop in the EDL is sized so that 
its outputs are immune to radiation strikes, 
guaranteeing that even in the presence of an SET, \emph{Corr} 
and \emph{Err} can never both be high and the asynchronous control logic has clean inputs.


We also note that more recent EDL designs can be
adopted that reduce the overhead of error detection \cite{kim2016450mv,Hua:2016}. 
In particular,
\cite{Hua:2016} proposed implementing multi-bit transition-detectors that significantly reduce the 
amortized area and power overheads.

\subsection{Metastability Analysis}
\label{sec:metastability}

There are two sources of metastability that must be considered in this design, in the error-detection logic and in the data path latches. 
First, metastability can occur in the EDL logic because of an internal SEU in the C-element or a race between its inputs. Fortunately, the Q-flop contains a special metastability filter~\cite{Rosenberger:1988} which guarantees that the Q-Flop outputs will remain zero until any internal metastability is resolved. 
Secondly, metastability can occur in the datapath. Under normal operation, the circuit is timed to ensure that the inputs of the latches satisfy the setup and hold time that is relative to the falling edge of the CLK. 
However, \revSecond{an} SET may violate this condition causing a latch to go into metastability. For this reason the SFW is set to the time period for which the clock is high {\em plus} the hold time of the latch. Consequently, any SET that causes a latch to go into metastability will be detected and mitigated by re-opening and closing the latch, essentially re-sampling the data.  This is in contrast to its timing-resilient
inspiration \cite{Peter:2015} which, relying on worst-case timing analysis, does not re-sample the data but instead mitigates timing violations only by slowing down downstream pipeline stages.  

Notice also that the circuitry that detects \revSecond{an} SET can invariably experience metastability if the SET occurs at the very beginning or end of the SFW. The benefit of the Q-flop is that it isolates the controllers from the metastability and, on average, resolves any metastability very quickly. The SERAD controller simply stalls until one of the dual-rail error signals \revSecond{is} asserted (which happens only after \rev{metastability} is resolved). In contrast, the only approach to resolve this synchronously that we are aware of requires waiting a worst-case delay of one to two clock cycles for metastability to resolve with a reasonable mean-time-between-failures. Unfortunately, one cannot decide whether to stall a pipeline one to two clock cycles after the potential metastability event occurs. For this reason most timing resilient synchronous designs either have ignored metastability \cite{Fojtik:2012Ra}, resulting in poor mean-time-between-failure \cite{beer2014metastability}, or have relied on flush-and-replay logic in which the entire pipeline is flushed of data and replayed (e.g., \cite{Das:2009}). Consequently, although some synchronous timing resilient schemes are also resilient to some SETs (as suggested in, for example, \cite{ARM}), none provide as comprehensive resilience as SERAD.

\begin{figure*}[tbh]
        \centering
        \includegraphics[width=\linewidth]{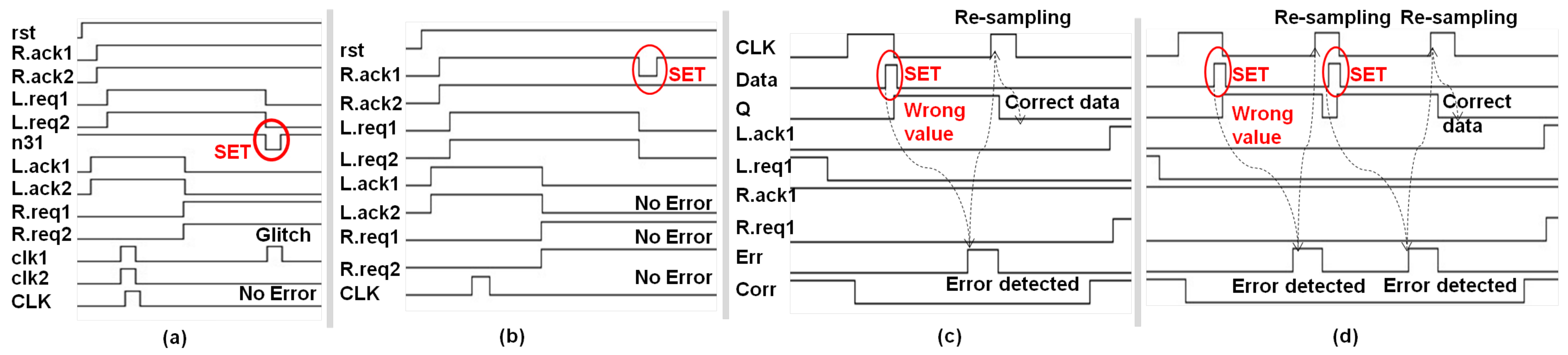}
        \caption{Verilog simulations showing (a) the resiliency of SERAD controllers to SETs; (b) SETs at the dual rail input of controllers; (c) the working of SERAD during SETs; (d) the working of SERAD during consecutive SETs.}
    \label{fig:sim}
    \vspace{-1em}
\end{figure*}

\subsection{Timing Constraints}

\label{sec:timing}
This section explains the timing constraints associated with SERAD. 
To simplify this discussion we denote
the delay components of the error detecting logic as follows.
\begin{itemize}
\item The propagation delay of the XOR gate i.e., from \emph{D} to \emph{X} of the EDL, is denoted $\delta_{xor-pd}$.
\item The output pulse width of the XOR gate is denoted $\delta_{X-pw}$.
\item The C-element's rising and falling propagation delays are denoted $\delta_{C-pullup}$ and $\delta_{C-pulldown}$, respectively. For simplicity, we assume these delays are larger than the C-element's minimum pulse width requirements.
\item The propagation delay of the OR tree between the C-elements and Q-Flop is denoted $\delta_{or-tree}$.
\item The Q-Flop's setup and hold times are denoted $\delta_{Q-setup}$ and $\delta_{Q-hold}$, respectively.
\end{itemize}

The output pulse of XOR gate should be sufficient wide so that it is captured by the C-element. This constraint can be approximated as
\begin{equation}
\delta_{X-pw} \ge \delta_{C-pullup}
\end{equation}
As illustrated in Fig.~\ref{fig:edl}, a 
small compensation delay $\delta_{comp}$ exists between 
the \emph{CLK} and \emph{Y} signals. 
It is implemented with an asymmetric delay line for which
$\delta_{comp-r}$ denotes the delay after a rising input and $\delta_{comp-f}$ denotes the delay after a falling input. Using this delay line, the path delays from \emph{CLK} to the rising transition of \emph{Y} and the delay from data to \emph{X} are matched. 
This ensures that the C-element registers the change in \emph{Data} as soon as the associated 
latches go transparent and remains sensitized until after the hold time of the latches expires. Assuming an ideal clock-tree, the constraints the delay line must adhere to are as follows.
\begin{equation}
\delta_{comp-r} = \delta_{xor-pd} + \delta_{X-pw}
\end{equation}
\begin{equation}
\delta_{comp-f} = \delta_{xor-pd} + \delta_{X-pw} + \delta_{DICE-hold}
\end{equation}
The Q-flop also has setup and hold constraints that should be observed (in the absence of metastability) to optimize performance.
\subsubsection{Setup Constraint} The Q-flop samples its input value when \emph{Sample} goes high. 
As discussed earlier, to properly catch all desired SETs, \emph{Sample} should go high only after the completion of \revThird{the} hold period of \revThird{the latches}, i.e. when \emph{Y} goes low, and the error signal propagates to the Q-flop inputs.
\begin{equation}
\delta_{Su} \ge \delta_{C-pullup} + \delta_{or-tree} + \delta_{Q-setup}
\end{equation}
$\delta_{Su}$ is the lower bound of the 
falling delay of the delay line between \emph{Sample} 
and \emph{Y}.
\subsubsection{Hold Constraint} The Q-flop input should be stable for its own hold time after the positive edge of sample. 
\begin{equation}
\delta_{Q-hold} \le \delta_{H} + \delta_{C-pulldown} + \delta_{or-tree}, 
\end{equation}
where $\delta_{H}$ is the minimum rising delay of the delay 
line between \emph{Sample} and \emph{Sample\_bar}.
Fortunately, this constraint is easily met given reasonable physical design. It is thus not required during synthesis but must be verified during the final post-physical-design sign-off procedure.

The EDL and controller delays also impose a lower bound on the non-overlap period $\delta_{no}$ between the two clocks of consecutive pipeline stages, as follows:
\begin{equation}
    \delta_{no} \geq \delta_{comp-r} + \delta_{Su} + \delta_{Q-pd} + \delta_{ctrl},
\end{equation}
where $\delta_{Q-pd}$ is the \emph{Sample} to \emph{Corr} delay of the Q-flop and $\delta_{ctrl}$ is the control logic delay from the \emph{Corr} signal of one pipeline stage to the rising edge of the \emph{CLK} signal of the next pipeline stage. If the worst-case delay of the combinational logic $\Delta$ is smaller than $\delta_{no} + \sigma$, this control overhead dictates the minimum cycle time of the design.  Otherwise, the control overhead can be completely hidden by using a smaller delay line $\delta$ between pipeline stages. 


\section{Experimental Results}
\label{sec:exp_results}

\revSecond{
The proposed radiation-hardened design template was tested and evaluated using both digital and analog test benches.}
\revThird{
The digital test benches used the Verilog digital simulator QuestaSim to demonstrate correct logical behavior of the template despite the presence of SETs. For these simulations, digital SETs were forcibly injected in the design.}
\revThird{
Spice-like analog simulation within the Cadence Virtuoso tool suite were used to demonstrate the effectiveness of gate sizing and the impact
of SETs that occur in the EDL. For these analog simulations SETs were modeled using double exponential current pulses whose specification were obtained using TCAD simulations.}

\subsection{Verilog Simulations}

\textbf{Case I:} Fig.~\ref{fig:sim}(a) shows \revSecond{an} SET in the controller \revSecond{that} will not propagate to the final outputs. An internal node (\emph{n31}) has \revSecond{an} SET which caused the output of one rail \emph{clk1} to have an error but the final output \emph{CLK} is error-free because the error is filtered by the sized guard gate.

\textbf{Case II:} Fig.~\ref{fig:sim}(b) shows that \revSecond{an} SET at the input of the dual-rail controllers does not result in error at any of the final outputs. In particular, it shows \revSecond{that} \revSecond{an} SET at input \emph{R.ack1} does not cause an error in any of the outputs, \emph{CLK, R.req1, R.req2, L.ack1} and \emph{L.ack2}.

\textbf{Case III:} Fig.~\ref{fig:sim}(c) shows what happens when \revSecond{an} SET in the combinational logic propagates to the input of the latch when clock is high. The clock (CLK), \revSecond{which re-samples the data, is closed and reopened,} marked as ``re-sampling" in Fig.~\ref{fig:sim}(c). Fig.~\ref{fig:sim}(d) shows an interesting sub-case when two SETs happens at the input of latch in consecutive cycles. Here, SERAD re-samples the data twice before it samples the correct data.

\begin{figure*}[!h]
        \centering
        \includegraphics[width=\linewidth]{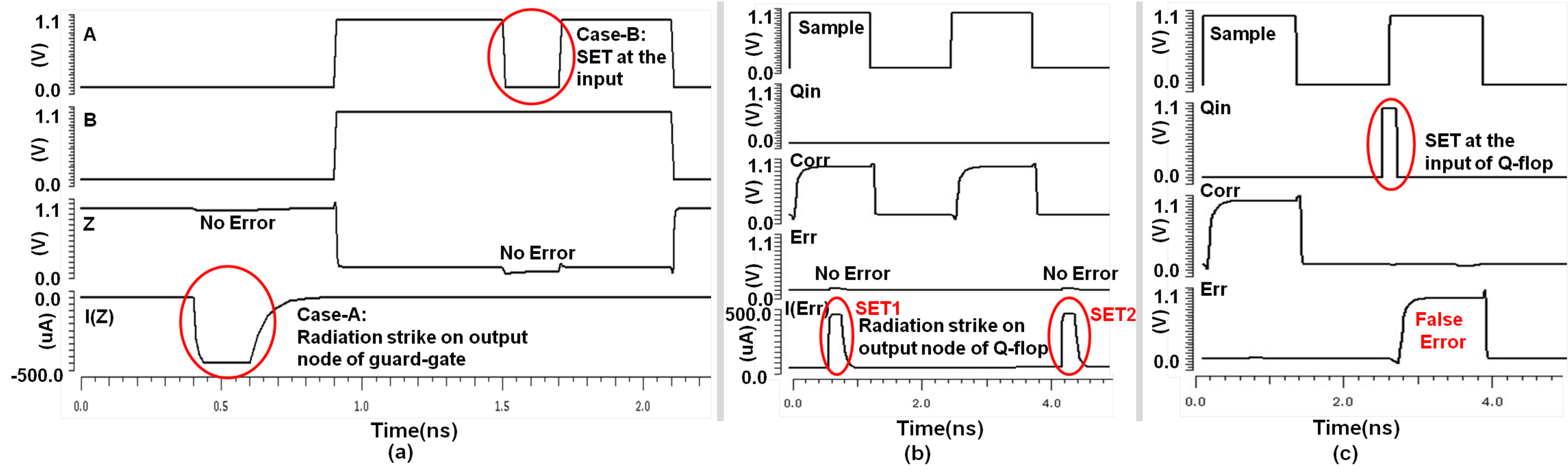}
        \caption{\rev{Spice simulations showing (a) resiliency of a sized guard-gate to \revThird{an SET} in the SERAD controller; (b) resiliency of a sized Q-flip when \revThird{a} particle strike occurs at one of the \revThird{outputs} of \revThird{a} Q-flop (\emph{Err});  (c) False error at the output of \revThird{a} sized Q-flop when \revThird{a} particle strike occurs at the input of the Q-flop.}}
    \label{fig:sim1}
    \vspace{-1em}
\end{figure*}

\subsection{Spice Simulations}

\textbf{Case IV:} A particle strike occurs at the controller output. \revThird{The} output does not show a glitch \revThird{because} the guard gates are \revThird{properly} sized. 
In particular, the simulations of the sized guard gate in Fig.~\ref{fig:sim1}(a) \revSecond{show} two cases: (i) Case-A shows the double-exponential SET current pulse~\cite{bala:2005} applied at the output of the sized guard-gate which does not propagate. (ii) Case-B indicates \revSecond{an} SET at the input of the guard-gate which also does not affect the output. 

\textbf{Case V:} Fig.~\ref{fig:sim1}(b) illustrates the case when a particle strike occurs at one of the \revSecond{outputs} of \revSecond{the} Q-flop (\emph{Err}). Since the Q-flop is \revThird{properly sized, the output of the Q-flop does not glitch.}  

\textbf{\rev{Case VI:}} Fig.~\ref{fig:sim1}(c) illustrates \revSecond{that} a soft-error in the EDL logic can be latched by Q-flop and lead to a false error. \revThird{This happens when an SET occurs in the error detecting logic that is detected by the Q-flop which consequently asserts its \emph{Err} signal. The controller conservatively interprets this as an indication of an SET in the combinational logic.} 
The controller \revThird{thus unnecessarily} re-samples ``stable" data.

We have also simulated the SERAD design under PVT variations, consisting of three process corners (FF, SS, TT), 10\% voltage variations, and temperature ranging from $-40^\circ C$ to $+70^\circ C$. We found process variations causes controller delay to vary by about 8\% around TT and that across all corners SET pulse widths vary by about 2.6\%. We observed that the SERAD controllers and EDL have sufficient margins to remain effective against \revThird{SETs} even under PVT variations.

\subsection{Plasma Case Study}

\revSecond{We also functionally validated the proposed template within an application in a radhard processor. In particular, we used}
a 3-stage Plasma~\cite{Plasma} processor, an OpenCore MIPS CPU, synthesized using a leading commercial synthesis tool 
and the NCSU 45nm open-source cell library. 
We compare our post-synthesized SERAD Plasma processor with three other variants:  
\begin{itemize}
    \item Sync Plasma: An unhardened synchronous implementation of Plasma processor used as a comparison baseline.
    \item Glitch-Filter Plasma: A radiation-hardened synchronous implementation of Plasma where glitch-filtering~\cite{bala:2005} is used to mitigate glitches in the datapath. This technique has less area and power overhead than triple modular redundancy 
    at the expense of a performance penalty 
    equal to twice the delay of a worst-case SET pulse.
    \item TMR Plasma:  A radiation-hardened synchronous implementation of Plasma  using the conventional technique of triple modular redundancy~\cite{TMR}. This technique incurs large area and power overheads but does not incur a significant performance penalty.
\end{itemize}
Power consumption is calculated at 286 MHz (the  maximum frequency achievable by all four compared variants) using the signal activity obtained from running the 
``count" \rev{and ``pi"} programs that are included in the Plasma open-source distribution.\footnote{\rev{The ``count" program performs $\sum_{i=1}^\infty 3^i$ and displays each term in the series in its word form and the ``pi" program numerically computes the value of $\pi$.} The ``Coremark" program contains implementations of the following algorithms: list processing (find and sort), matrix manipulation (common matrix operations), state machine processing to determine if an input stream contains valid numbers), and a CRC (cyclic redundancy check).
}

More specifically, a combination of Python and TCL scripts were used to convert the original synchronous design into the radiation hardened variants. Custom cells were made for the guard gate and DICE \revSecond{FFs} and latches. The glitch filter version requires an SET filter~\cite{bala:2005} at 
the input of each \revSecond{FF} and all \revSecond{FFs} are replaced with DICE \revSecond{FFs} with SET filters in this design. 
For the TMR version, the combinational logic and FFs are triplicated and voters are added to each triplet of FFs.

Similarly, the synchronous Plasma design is converted into a SERAD design using a semi-automated CAD flow similar to the \rev{timing-resilient so-called "Blade"} flow presented in~\cite{Peter:2015}. 
\rev{The Blade flow involves automatically synthesizing the RTL to gates using standard \revSecond{FFs} and using a custom TCL script to replace the FFs with two-phase master-slave latches, re-time the slave latches, replace timing-critical latches with error-detecting latches, and automatically add the EDL 
and Blade controllers to manage the local clocks. Our SERAD flow is different in 
that all latches are replaced with DICE latches (rather than only the timing-critical latches) and that the added EDL and SERAD controllers (described in Sections \ref{sec:controller} \& \ref{sec:EDL}) are different than those used for Blade \cite{Peter:2015}.}
\rev{As with the Blade flow, the SERAD} flow is applicable to any RTL synchronous specification\rev{.}

Table~\ref{tab:table1} 
shows the \rev{\emph{maximum}} performance 
\rev{and associated area} 
\rev{of} the four variants of Plasma using the synchronous design as a baseline.
According to Table~\ref{tab:table1}, the area of SERAD Plasma design is 80\% higher than the baseline but comparable to the Glitch-Filter Plasma and less than half of the TMR design. This is because the added relative cost of the error detecting logic (EDL) and control logic are not large, and the total cost of DICE latches is not significantly more than that of the DICE FFs (despite having two-times more latches than FFs). 
Notice that the Glitch-Filter Plasma has the highest performance degradation due to the additional delay of twice of maximum SET pulse width ~\cite{Hosseinabady:2006}. It is important to emphasize that this cost is fixed and thus is more prominent for high-frequency designs. 

\begin{table}[h]
\scriptsize\addtolength{\tabcolsep}{-0.3pt}
\centering
  \begin{tabular}{|l|c|c|c|c|c|}
    \hline
    \multirow{2}{*}{Design type} & Max. & \multicolumn{4}{|c|}{Area}\\ \cline{3-6}
    & Freq. & Comb & Seq & Total & Incr. (\%)\\ \hline 
    \bfseries{SERAD Plasma} & \bfseries{333} & \bfseries{55784} & \bfseries{25000} & \bfseries{80784} &  \bfseries{80.6}\\ \hline
    Sync Plasma~\cite{Plasma} &  340 & 33083 & 11656 & 44739 & 0\\  \hline
    Glitch-Filter Plasma~\cite{bala:2005} &  286 & 57867 & 23439 & 81306 & 81.7\\ \hline
    TMR Plasma~\cite{TMR} & 329 & 143679 & 34969 & 178648 & 299 \\ \hline
  \end{tabular}
  \caption{Maximum frequency (MHz) and area ($\mu m^2$) comparison of the four design variants}
  \label{tab:table1}
\end{table}
\begin{table}[h]
\scriptsize\addtolength{\tabcolsep}{3pt}
\centering
  \begin{tabular}{|l|c|c|c|c|}
    \hline
    Design type & Comb & Seq & Total & Incr. (\%)\\
    \hline 
     \bfseries{SERAD Plasma} & \bfseries{2.57} & \bfseries{4.17} & \bfseries{6.74} & \bfseries{-23.7}\\ \hline
    Sync Plasma~\cite{Plasma} & 2.12 &  6.71 & 8.83 & 0 \\  \hline
    Glitch-Filter Plasma~\cite{bala:2005} & 2.97 &  10.2 & 13.2 & 49.5 \\ \hline
    TMR Plasma~\cite{TMR}& 7.69 & 18.34 & 26 & 194.4 \\ \hline
 \end{tabular}
 \caption{Power (mW) comparison of the four design variants \revSecond{with a 286MHz clock running} the ``count" program}
  \label{tab:table2}
\end{table}
\begin{table}[h]
\scriptsize\addtolength{\tabcolsep}{3pt}
\centering
  \begin{tabular}{|l|c|c|c|c|}
    \hline
    Design type & Comb & Seq & Total & Incr. (\%)\\
    \hline 
     \bfseries{SERAD Plasma} & \bfseries{2.54} & \bfseries{4.82} & \bfseries{7.36} & \bfseries{-25.1}\\ \hline
    Sync Plasma~\cite{Plasma} & 2.93 &  6.88 & 9.82 & 0 \\  \hline
    Glitch-Filter Plasma~\cite{bala:2005} & 4.91 & 10.46 & 15.37 & 56.5 \\ \hline
    TMR Plasma~\cite{TMR}& 11.26 & 18.64 & 29.9 & 204.5 \\ \hline
 \end{tabular}
 \caption{Power (mW) comparison of the four design variants \revSecond{with a 286MHz clock running} the ``pi" program}
  \label{tab:table3}
\end{table}
\begin{table}[h]
\scriptsize\addtolength{\tabcolsep}{3pt}
\centering
\begin{tabular}{|l|c|c|c|c|}
    \hline
    Design type & Comb & Seq & Total & Incr. (\%)\\
    \hline 
     \bfseries{SERAD Plasma} & \bfseries{2.59} & \bfseries{4.74} & \bfseries{7.34} & \bfseries{-12.8}\\ \hline
    Sync Plasma~\cite{Plasma} & 2.26 & 6.16  & 8.42 & 0 \\  \hline
    Glitch-Filter Plasma~\cite{bala:2005} & 3.05 & 6.58 & 9.63 & 14.3 \\ \hline
    TMR Plasma~\cite{TMR}& 8.26 & 18.43 & 26.69 & 216.9 \\ \hline
 \end{tabular}
 \caption{\revSecond{Power (mW) comparison of the four design variants with a 286MHz clock running the ``CoreMark" program}}
  \label{tab:table4}
\end{table}

\rev{Table~\ref{tab:table2}, Table~\ref{tab:table3}}, \revSecond{and Table~\ref{tab:table4}} \revSecond{compare} the power consumption among the four design variants at the clock frequency of 286MHz using ``count," ``pi", \revSecond{and ``CoreMark"} programs, 
respectively.\footnote{\revSecond{Because the four variants compute exactly the same operation across every cycle, running all variants at the same frequency also provides a direct and fair comparison of energy consumption.}}
Interestingly, the SERAD design is comparable to unhardened baseline in terms of performance
and, somewhat counter-intuitively, is estimated to consume 
lower power. 
In particular, as detailed in \revSecond{the three power comparison tables, SERAD consumes relatively
low sequential power.} This is a result of three factors.
First, compared to the default FF design in the unhardened version, there is no local buffer in a DICE latch, which reduces clock switching power. The impact of this is particularly significant in this case study because of the relatively low switching 
activity on the data signals. Interestingly, this type of cell design has been independently proposed in error-detecting latch-based designs because the removed local clock buffers can be more efficiently compensated during physical design by properly adjusting the clock tree \cite{kim2016450mv}.
Second, the input capacitance of the clock pin in the default FF is larger than in the DICE latch. In our library, the input capacitance is 8fF for the FF versus 5fF for the DICE latch.
Lastly, the latch-based design yields lower switching activity at the sequential elements by eliminating unnecessary glitches. For our particular experiment, the data pin of the DICE latches has an average activity factor of 2.0\% compared to \revSecond{3.7\%} for the data pin of FFs in the unhardened design.
\revSecond{In addition, we also note that the increase in area is due largely to the error detection logic which in normal operation does not switch.} This further explains why despite being larger, we see lower power consumption than the other three variants (operating at the same frequency). 
However, we do see about $5.9\times$ increase in the leakage power as compared to sync Plasma.
\revSecond{In particular, the fraction of total power due to leakage in our SERAD design grew to 22.6\%.}
This is due to the fact we are significant larger in size and because our DICE latches have not been optimized for leakage. 

While this power analysis comparison is promising, it is also important to recognize that these results do not take into account the power of the clock tree that is designed during and accounted for after physical design. In particular, two-phase latch-based designs require two clock trees which may present an additional power overhead. Such latch-based designs, however, have built-in hold margins and thus require fewer hold buffers (also added during physical design) than their FF-based counterparts. 
The combined impact of these effects 
is process, cell-library, and design dependent.  
\rev{
Given efficient clock tree synthesis for asynchronous designs is an on-going research and engineering challenge (see e.g., \cite{gimenez2018static}) we have estimated this overhead by place-and-routing synchronous \revSecond{FF} and two-phased latch-based Plasma designs and found the clock tree power increased by 12\% representing approximately a 2\% increase in overall power.} 

Lastly, it is important to note that these results do not take into account the ability of the asynchronous design to track process, voltage, and temperature variations. Because the delay of the control circuits is generally positively correlated with the delay of the combinational logic, smaller margins are needed in the delay lines than in synchronous clock periods. Although difficult to quantify, this leads to increased performance and/or increased yield \cite{Cortadella:2016,Zhang:2017}, the degree of 
which is dependent on the relative amount of local versus global variation, whether speed binning is employed, and whether chips are allowed to vary in performance due to transient voltage or temperature variations.



\section{Conclusions}
\label{sec:conclusions}

This paper presents a design template for soft-error resilient asynchronous bundled-data design called SERAD that uses a novel combination of space and temporal redundancy to become resilient to SETs. 
The SERAD design template has been validated in a 45nm technology using a combination of Spice and Verilog simulations. The resulting area, performance, and power of the design template has been evaluated 
on an open-core MIPS-like processor. Compared with the unhardened synchronous version, the post-synthesis SERAD design is 81\% larger, exhibits negligible performance degradation, and is estimated to consume lower power. It consumes less than half of the area of the TMR design and is significantly faster than the glitch-filtering-based design, making it a promising approach for radiation hardening.

There are two areas of future work that can improve the benefits of SERAD and expand its applicability. 
First, we can improve the performance of SERAD by leveraging its inherent timing-resilience by modifying its control to also recover from uncommonly long combinational delays. \revSecond{In particular, if the latch transparency 
window is started early, near-critical combinational delays will be detected
as errors. They will trigger the 
error detecting logic and be
mitigated via re-sampling.}  
This should cost negligible area but lead to significant further performance improvements over unhardened synchronous designs. Second, we propose to design radiation-hardened asynchronous-synchronous clock domain crossing modules that can surround a SERAD design, enabling it as a drop-in replacement for latency-insensitive synchronous modules.

\revSecond{
\section*{Acknowledgements}
We would like to acknowledge the help of Dr. Dylan Hand in porting the CoreMark program to Plasma. This work is supported in part by a grant received from the Ministry of Electronics and Information Technology (MEITY), Government of India for a Special Manpower Development Project for Chips to System Design (SMDP-C2SD) and by a grant received from the Science and Engineering research board (SERB) grant CRG/2018/005013  .
}

\begin{IEEEbiography}
[{\includegraphics[width=1in, height=1.25in, clip, keepaspectratio]{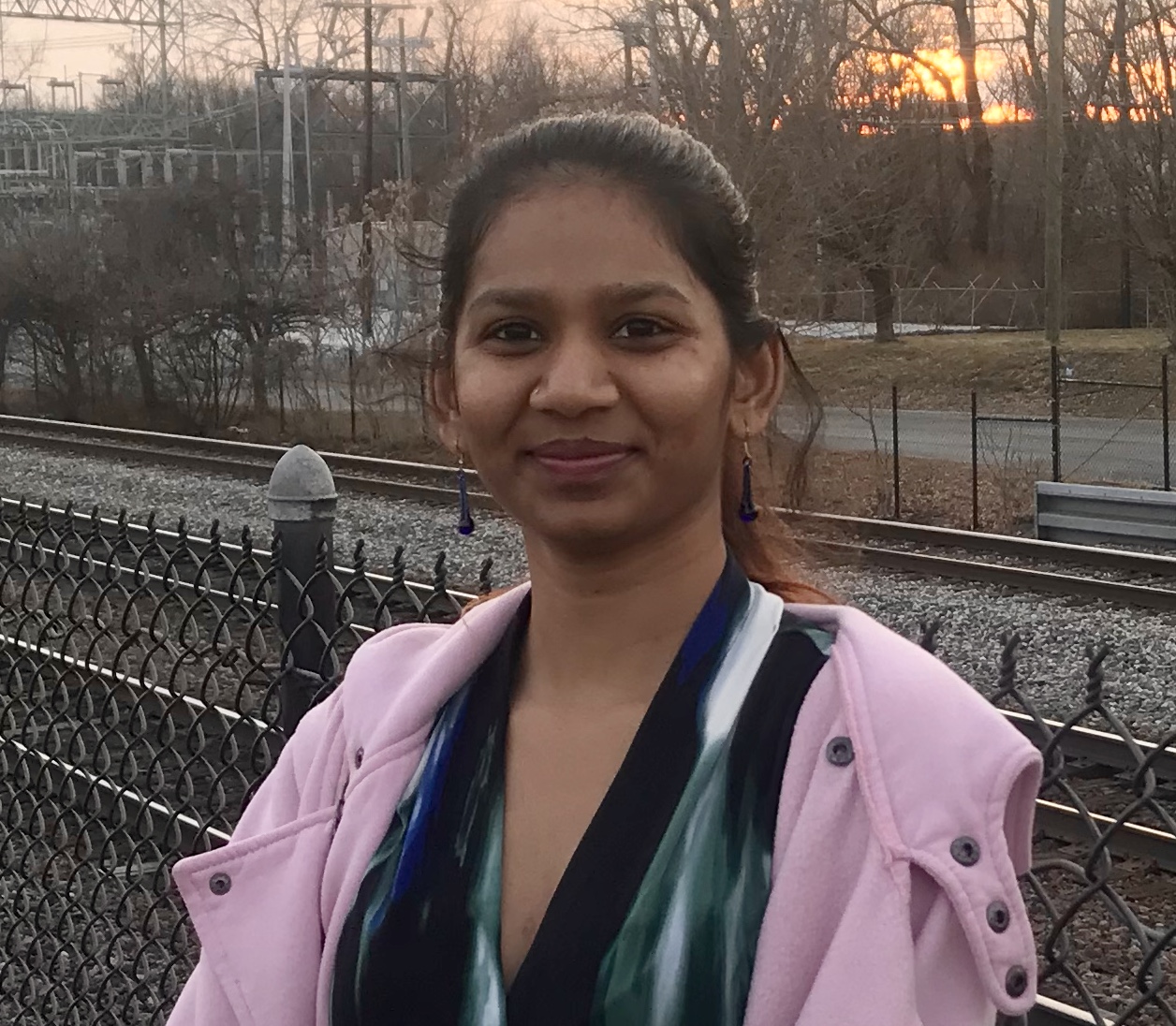}}]
{\textbf{Sai Aparna Aketi}} received her B.Tech degree in Electrical Engineering at Indian Institute of Technology Gandhinagar in 2018. This work was started when Aparna Aketi was interning at the University of Southern California as part of IUSSTF-Viterbi Program in 2017. She is currently pursing her doctoral degree under the guidance of Prof. Kaushik Roy at Centre for Brain Inspired Computing (C-BRIC), Purdue University. Her current research interests include a variety of topics in explainable, robust and energy-efficient deep learning.
\end{IEEEbiography}
\vspace{-1.5em}
\begin{IEEEbiography}
[{\includegraphics[width=1in, height=1.25in, clip, keepaspectratio]{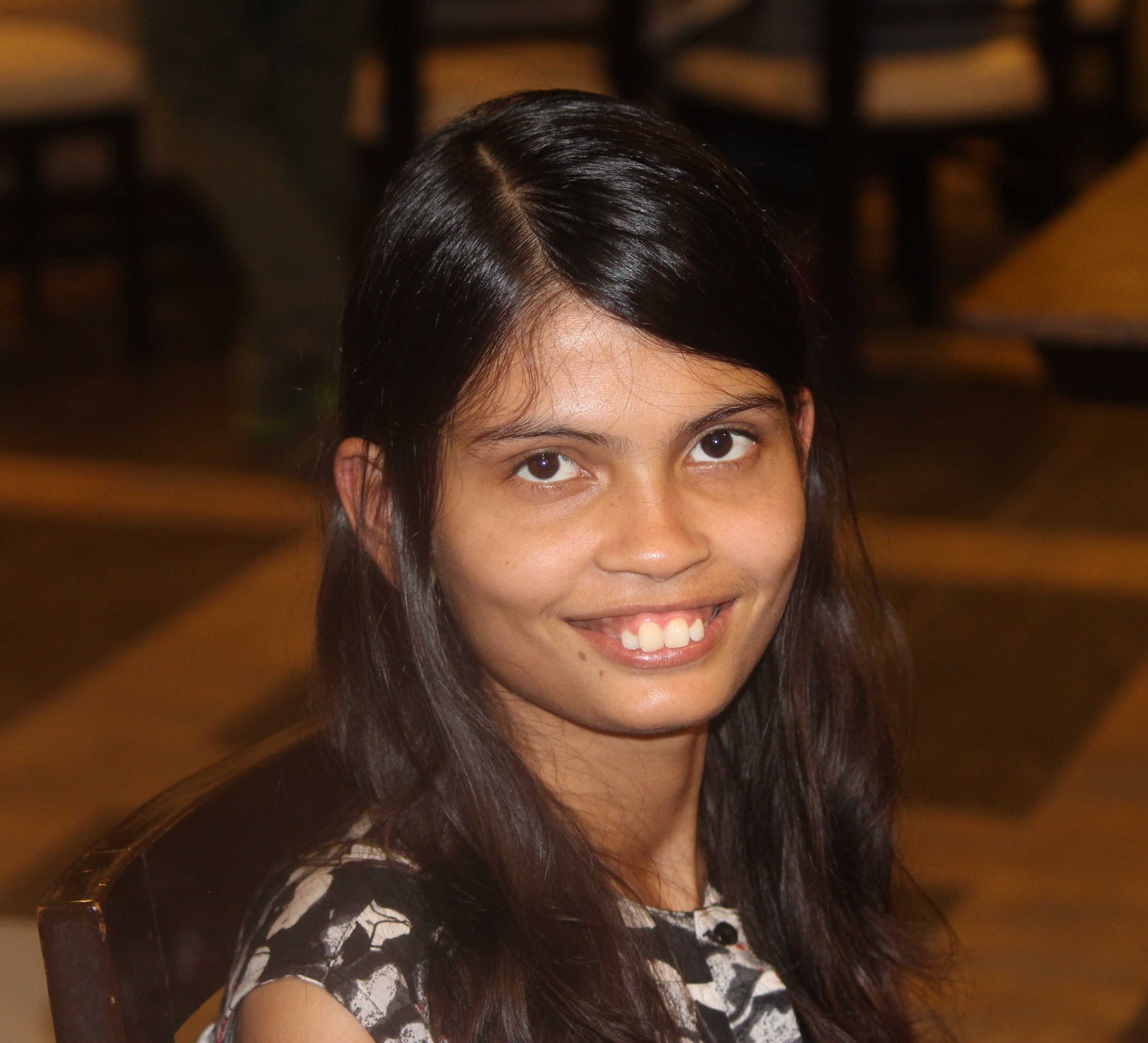}}]
{\textbf{Smriti Gupta}} received her B.Tech degree in Electronics \& Communication Engineering from Institute of Engineering \& Technology Lucknow, India in 2016 and her M.Tech degree in Electrical Engineering from Indian Institute of Technology Gandhinagar in 2018. She is currently working as a Senior Engineer in MediaTek Bangalore, India. Her interests include high performance and low power VLSI designs.
\end{IEEEbiography}
\vspace{-1.5em}
\begin{IEEEbiography}
[{\includegraphics[width=1in, height=1.25in, clip, keepaspectratio]{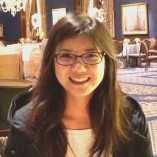}}]
{\textbf{Huimei Cheng}}
Huimei Cheng is a Ph.D. student in Ming Hsieh Department of Electrical and Computer Engineering at the University of Southern California. She received her B.S. degree at Nanjing University of Information \& Technology (China) in 2014, and her M.S. degree from USC in 2016. Upon graduation, she worked at Synopsys in R\&D Prime Time team conducting research on pessimism reduction in crosstalk. She is a student member of IEEE. Her research interests include a variety of topics in CAD and asynchronous VLSI design.
\end{IEEEbiography}
\vspace{-1.5em}
\begin{IEEEbiography}
[{\includegraphics[width=1in, height=1.25in, clip, keepaspectratio]{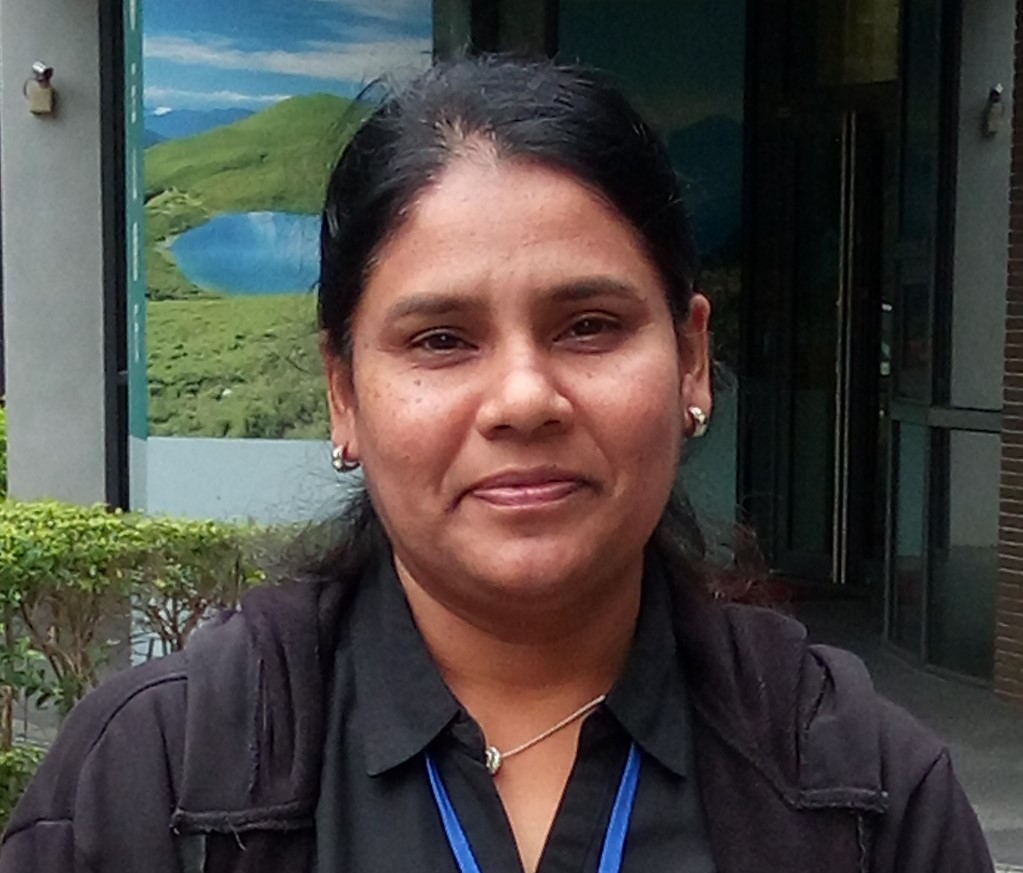}}]
{\textbf{Joycee Mekie}} received her Ph.D. degreee in Electrical Engineering from IIT Bombay in 2009, and received her Bachelors and Masters in Electrical Engineering from M. S. University of Baroda in 1997 and 1999, respectively. She joined as Assistant Professor in the Electrical Engineering Department at IIT Gandhinagar in 2009. She is a recipient of the prestigious Young Faculty Research Fellowship (YFRF) from Ministry of Electronics and Information Technology under the Visvesvaraya PhD scheme. She has served on the technical program committee of several conferences, including ASYNC and VLSID, and is the reviewer for several journals, including TCASI, TCASII and TCAD. Her research interests include Approximate computing, Circuits for space applications, Asynchronous systems, Energy-efficient memory design, Computer architecture and Network-on-chip architectures.
\end{IEEEbiography}
\vspace{-1.5em}
\begin{IEEEbiography}
[{\includegraphics[width=1in, height=1.25in, clip, keepaspectratio]{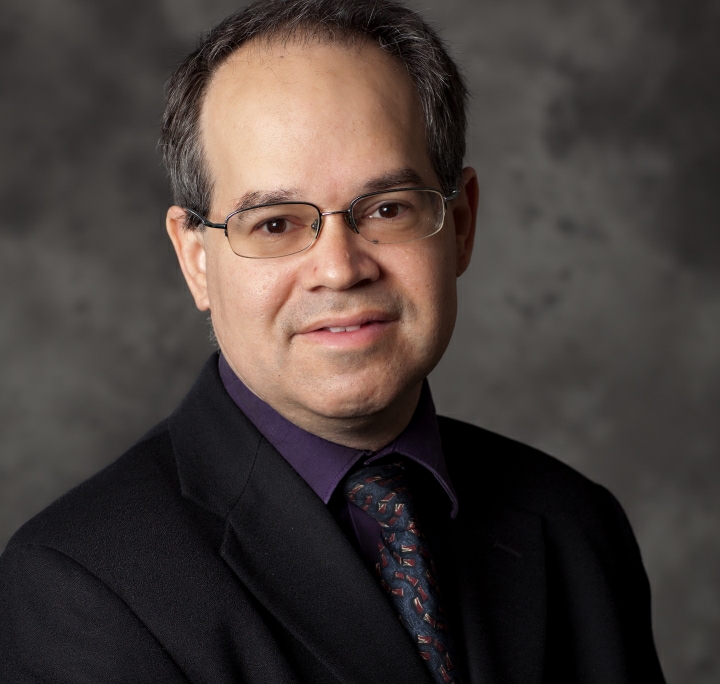}}]
{\textbf{Peter A. Beerel}}
received his B.S.E. degree in Electrical Engineering from Princeton University, Princeton, NJ, in 1989 and his M.S. and Ph.D. degrees in Electrical Engineering from Stanford University, Stanford, CA, in 1991 and 1994, respectively. Professor Beerel is currently a Full Professor and Associate Chair of the Computer Engineering Division of the Ming Hsieh Electrical and Computer Engineering Department at the University of Southern California. He co-founded TimeLess Design Automation to commercialize an asynchronous ASIC flow in 2008 and sold the company in 2010 to Fulcrum Microsystems which was bought by Intel in 2011. His interests include a variety of topics in CAD, VLSI, and Machine Learning. He is a Senior Member of the IEEE.
\end{IEEEbiography}
\end{document}